\documentclass[prc,preprintnumbers,twocolumn]{revtex4}
\usepackage{graphicx}
\usepackage{amssymb}
\usepackage{amsmath}

\begin{document}

\title{Quasifission at extreme sub-barrier energies}
\author{V.V.Sargsyan$^{1,2}$, G.G.Adamian$^{1}$, N.V.Antonenko$^1$, W. Scheid$^3$, and  H.Q.Zhang$^4$
}
\affiliation{$^{1}$Joint Institute for Nuclear Research, 141980 Dubna, Russia\\
$^{2}$International Center for Advanced Studies, Yerevan State University, M. Manougian 1, 0025, Yerevan, Armenia\\
$^{3}$Institut f\"ur Theoretische Physik der Justus--Liebig--Universit\"at, D--35392 Giessen, Germany\\
$^{4}$China Institute of Atomic Energy, Post Office Box 275, Beijing 102413,  China
}
\date{\today}

\begin{abstract}
With the quantum diffusion approach the  behavior of the capture cross-section
is investigated in the reactions $^{92,94}$Mo + $^{92,94}$Mo, $^{100}$Ru + $^{100}$Ru, $^{104}$Pd + $^{104}$Pd,
and $^{78}$Kr + $^{112}$Sn at deep sub-barrier energies
which are lower than the ground state energies of the compound
nuclei. Because the capture cross section
is the sum of the complete fusion and quasifission cross sections,
and the complete fusion cross section is zero at these
sub-barrier energies, one can study experimentally the
unique  quasifission process in these reactions after the capture.

\end{abstract}

\pacs{25.70.Jj, 24.10.-i, 24.60.-k \\ Key words: sub-barrier capture,
quantum diffusion approach, quasifission}

 \maketitle

%\section{Introduction}
The first evidences of hindrance  for compound nucleus
formation in the reactions with massive nuclei ($Z_1\times Z_2>1600$)
at  energies near the Coulomb barrier
were observed at GSI already long time ago~\cite{GSI,GSI2,GSI3}.
The theoretical investigations showed that the probability of complete fusion  depends
on the competition between the complete fusion and quasifission
after the capture stage~\cite{Volkov,nasha,Avaz}. As known, this competition can strongly reduce
the value of the fusion cross section and, respectively,
the value of the evaporation residue cross section
in reactions producing heavy and superheavy nuclei.
The quasifission is related to the binary decay of the nuclear system after the capture, but before a compound nucleus
is formed which could exist at angular momenta treated~\cite{Volkov,Schroder,nasha,Avaz}.
The quasifission process was originally ascribed only to reactions with massive  nuclei. But it is the general
phenomenon which takes place in reactions with the massive and
medium-mass  nuclei at energies above and below the Coulomb barrier~\cite{EPJSub1,EPJSub2}.
The mass and angular distributions of the
quasifission products depend on the entrance channel and the bombarding energy~\cite{Schroder}.
%Because the capture cross section is the sum of the fusion and quasifission cross sections,
%from the comparison of calculated capture cross sections and measured
%fusion cross sections one can extract the hindrance factor and
%show a  role of the quasifission channel
%in the reactions with various medium-mass and heavy nuclei at extreme
%sub-barrier energies.

For systems with negative $Q$-value,
the complete fusion cross section $\sigma_{fus}$ is equal to zero at
bombarding energies $E_{\rm c.m.}<E_{\rm c.m.}^{0}=-Q$:
$$\sigma_{fus}(E_{\rm c.m.}<E_{\rm c.m.}^{0})=0.$$
This expression implies  that the fusion cross
section or the fusion probability $P_{fus}$
must go to zero when the center-of-mass energy $E_{\rm c.m.}$
approaches the ground state energy, -$Q$, of the compound
nucleus.
Since the sum of the complete fusion cross section $\sigma_{fus}$
and the quasifission cross section $\sigma_{qf}$
gives the capture cross section
$$\sigma_{cap}=\sigma_{fus}+\sigma_{qf},$$
at $E_{\rm c.m.}<E_{\rm c.m.}^{0}=-Q$
we have
$$\sigma_{cap}(E_{\rm c.m.}<E_{\rm c.m.}^{0})=\sigma_{qf}.$$
So, at these deep sub-barrier energies the quasifission is
only contribution to the capture cross section and there is no the overlapping
between the fusion-fission and quasifission processes
as at higher bombarding energies.
%The hindrance factor which is equal to 1 may be understood in term of quasifission.
At deep sub-barrier energies,
the quasifission event  corresponds to the formation
of a nuclear-molecular state or dinuclear system
with small excitation energy that separates
by quantum tunneling through the Coulomb barrier
 in a binary event with mass and charge  close
to the entrance channel.
%In this sense the quasifission is the general phenomenon
%which takes place in the reactions with the massive,
%medium-mass  and,  probably, light nuclei~\cite{Volkov,nasha,Avaz,GSI,EPJSub1,EPJSub2}.

Although many measurements do not reach such deep sub-barrier energies $E_{\rm c.m.}<E_{\rm c.m.}^{0}=-Q$,
it is still possible to find systems with relatively small values of
$V_b - E_{\rm c.m.}^{0}=V_b + Q$ ($V_b=V(R_b)$ is the height of the  Coulomb barrier for the
spherical nuclei, $R_b$ is the position of this barrier) for the experimental  study of the quasifission process.
The purpose of the present article is to find such type of systems and
to estimate the capture cross sections at $E_{\rm c.m.}<E_{\rm c.m.}^{0}=-Q$.
The  quantum diffusion approach~\cite{EPJSub1,EPJSub2,EPJSub,EPJSub3}
is applied to study the capture  process more thoroughly.

%\section{Model}
In our quantum diffusion approach~\cite{EPJSub1,EPJSub2,EPJSub,EPJSub3}
the collisions of  nuclei are treated in terms
of a single collective variable: the relative distance  between
the colliding nuclei. The  nuclear deformations are taken into account through the dependence
of the nucleus-nucleus potential on the quadrupole deformations and mutual  orientations of the colliding nuclei.
Our approach regards the fluctuation and dissipation effects in
the collision of heavy ions and models the coupling with various channels
(for example, coupling of the relative motion with low-lying collective modes
such as dynamical quadrupole and octupole modes of the target and projectile~\cite{Ayik333}).
We have to mention that many quantum-mechanical and non-Markovian effects accompanying
the passage through the potential barrier are considered in our formalism~\cite{EPJSub,our}
through the friction and diffusion.
To calculate the nucleus-nucleus interaction potential $V(R)$,
we use the procedure presented in Refs.~\cite{EPJSub,EPJSub1,EPJSub2}.
For the nuclear part of the nucleus-nucleus potential, the double-folding formalism with
a Skyrme-type density-dependent effective nucleon-nucleon interaction is used.
%The parameters of the potential were adjusted to describe the experimental
%data at energies above the Coulomb barrier corresponding to spherical nuclei.
The absolute values of the quadrupole deformation parameters $\beta_2$
of deformed nuclei were taken from Ref.~\cite{Ram}.

%The two-neutron
%transfer with the  positive  $Q_{2n}$-value was taken into consideration~\cite{EPJSub,EPJSub2}.
The calculated results for all reactions
are obtained with the same set of parameters as in Refs.~\cite{EPJSub,EPJSub2}
and are rather insensitive to a reasonable variation of them.
One should stress that diffusion models, which  also include  quantum statistical effects,
were proposed in Refs.~\cite{Hofman,Ayik,Hupin} too.

%The heights of the calculated Coulomb barriers $V_b=V(R_b)$
%($R_b$ is the position of the Coulomb barrier)
%are adjusted to the experimental data
%for the fusion or capture cross sections.

%\section{Results of calculations}
Symmetric and near symmetric dinuclear systems with
neutron-deficient stable nuclei have the smallest values of $(V_b + Q)$.
For example, the sub-barrier energies with respect to the Coulomb barrier are
$V_b - E_{\rm c.m.}^{0}=V_b + Q=13, 14.8, 18, 19.4,  21.8$ MeV for the systems
$^{92}$Mo + $^{92}$Mo, $^{104}$Pd + $^{104}$Pd,  $^{94}$Mo + $^{94}$Mo,
$^{100}$Ru + $^{100}$Ru,  $^{78}$Kr + $^{112}$Sn, respectively.
Here predictions of  unknown mass-excesses of the compound nuclei are taken from Ref.~\cite{MN}.
In Figs.~1--3 the calculated capture cross sections for these reactions
are presented.
\begin{figure}
\vspace*{-0.cm}
\centering
\includegraphics[angle=0, width=0.8\columnwidth]{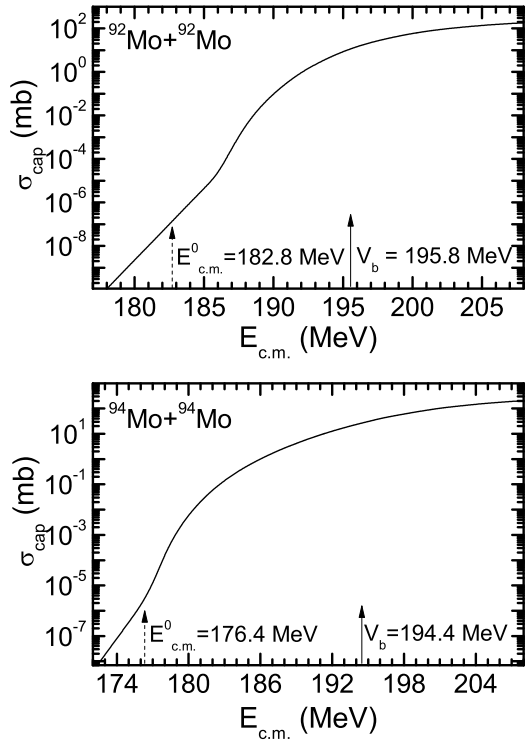}
\vspace*{-0.2cm}
\caption{The calculated capture cross sections vs $E_{\rm c.m.}$ for the reactions
$^{92,94}$Mo + $^{92,94}$Mo.
The dashed and solid arrows show  $E_{\rm c.m.}=E_{\rm c.m.}^{0}=-Q$
and   $E_{\rm c.m.}=V_b$, respectively.
}
\label{1_fig}
\end{figure}
\begin{figure}
\vspace*{-0.cm}
\centering
\includegraphics[angle=0, width=0.8\columnwidth]{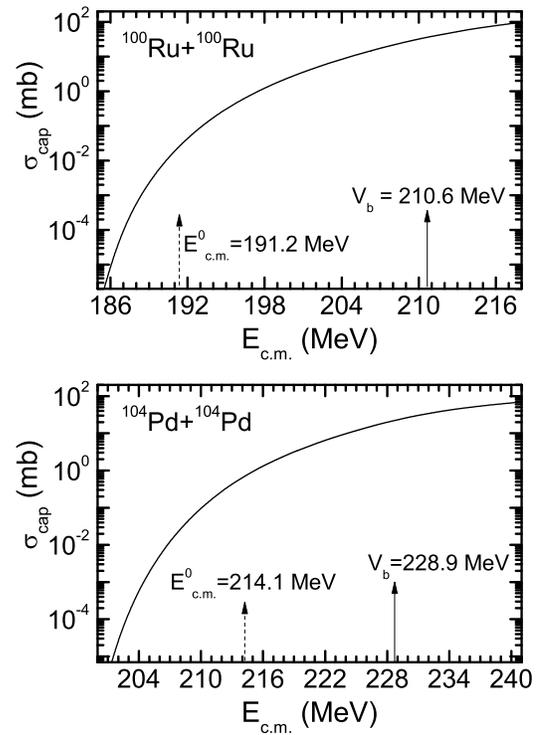}
\vspace*{-0.2cm}
\caption{The same as in Fig.~1, but for the  reactions $^{100}$Ru + $^{100}$Ru and $^{104}$Pd + $^{104}$Pd.
}
\label{2_fig}
\end{figure}
\begin{figure}
\vspace*{-0.cm}
\centering
\includegraphics[angle=0, width=0.8\columnwidth]{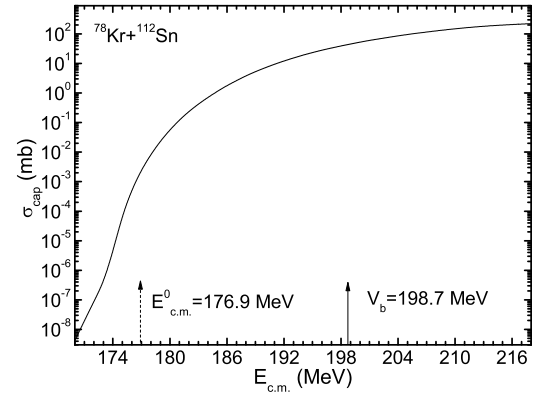}
\vspace*{-0.2cm}
\caption{The same as in Fig.~1, but for the $^{78}$Kr + $^{112}$Sn reaction.
}
\label{3_fig}
\end{figure}
All systems show a steady decrease of the sub-barrier fusion cross sections
with a pronounced change of slope.
%All reactions exhibit a steep fall-off in the capture cross section
%at  energies just under the barrier.
With $E_{\rm c.m.}$ decreasing   below
the Coulomb barrier the interaction changes because at the external
turning point the colliding nuclei do no more reach the region of the nuclear interaction
where the friction plays a role.
As result, at smaller $E_{\rm c.m.}$ the cross sections fall with a smaller rate.
For sub-barrier energies, the results of calculations
are very sensitive to the  quadrupole
deformation parameters $\beta_2$ of the interacting nuclei.
The influence of nuclear deformation is straightforward.
If the target and projectile nuclei are prolate in their ground states,
the Coulomb field on its tips is lower than on its sides. This
increases the capture  probability at  energies below
the barrier corresponding to the spherical nuclei.
The  enhancement of sub-barrier capture for the reactions
$^{104}$Pd + $^{104}$Pd,
$^{100}$Ru + $^{100}$Ru, and $^{78}$Kr + $^{112}$Sn in the contrast to the
reactions
$^{92,94}$Mo + $^{92,94}$Mo
is explained by the deformation  effect:
the deformations in the former systems are larger the ones in the later systems.

In Figs.~1--3 the calculated capture cross sections
at $E_{\rm c.m.}=E_{\rm c.m.}^{0}=-Q$ are $\sigma_{cap}=$0.2 nb,  5.1 nb, 2.3 $\mu$b, 24.4 $\mu$b,
and 0.7 mb for the reactions
$^{92}$Mo + $^{92}$Mo, $^{94}$Mo + $^{94}$Mo, $^{78}$Kr + $^{112}$Sn,
$^{100}$Ru + $^{100}$Ru,  and  $^{104}$Pd + $^{104}$Pd, respectively.
So, $^{104}$Pd + $^{104}$Pd, $^{100}$Ru + $^{100}$Ru, and $^{78}$Kr + $^{112}$Sn
are the optimal reactions for studying capture and quasifission at deep sub-barrier
energies $E_{\rm c.m.} < E_{\rm c.m.}^{0}=-Q$ where
the  complete fusion channel is closed ($\sigma_{fus}=0$).
At these sub-barrier energies the quasifission process can be studied in  future experiments:
from the measurement of the mass (charge) distribution
in collisions with total momentum transfer one can
show the distinct components which are due to quasifission
(with respect to the quasielastic components).
Because  the angular momentum is $J<10$ at these energies,
the angular distribution would have a small anisotropy.
The low-energy experimental quasifission data would probably provide straight information
since the high-energy data may be shaded by competing the
fusion-fission processes.
%More experimental and theoretical studies of sub-barrier fusion hindrance
%are needed to improve our understanding of pure quasifission process.
%One can try  to check experimentally these predictions.
The lifetime of nuclear molecule formed seems to be long enough to separate it
mass from other reaction products. Then one can observe the decay of this molecule into two fragments.

%\section{Summary}
In conclusion, the quantum diffusion approach
was applied to calculate the capture cross sections for the
reactions $^{92}$Mo + $^{92}$Mo, $^{104}$Pd + $^{104}$Pd,  $^{94}$Mo + $^{94}$Mo,
$^{100}$Ru + $^{100}$Ru, and $^{78}$Kr + $^{112}$Sn
at extreme sub-barrier energies
which are too low for complete fusion. The  quasifission near the entrance channel
is the unique binary decay process after the capture.
The reactions $^{104}$Pd + $^{104}$Pd, $^{100}$Ru + $^{100}$Ru, and $^{78}$Kr + $^{112}$Sn
seem to be optimal systems for a experimental study  of the true quasifission  at extreme sub-barrier energies.

%We thank H.~Jia, J.Q. Li, C.J.~Lin, and S.-G.~Zhou for fruitful discussions and  suggestions.
%We are grateful to J.F.~Liang for providing us his experimental data.
This work was supported by DFG, NSFC, and RFBR.
The IN2P3(France) - JINR(Dubna)
%, MTA(Hungary)-JINR(Dubna)
and Polish - JINR(Dubna)
Cooperation Programmes are gratefully acknowledged.\\

%\newpage

\end{document}